\documentclass[12pt]{article}
\usepackage{graphicx}% Include figure files
\newcommand{\sss}{\scriptscriptstyle}

\newcommand {\be}{\begin{equation}} % start equation
\newcommand{\ee}{\end{equation}}    % end equation

\def\dds1{\frac{\partial}{\partial s_1}}

\def\vtn{v_{{\sss T}n}}
\def\vti{v_{{\sss T}i}}
\def\vte{v_{{\sss T}e}}

\def\d{d\kern-0.8 ex\vrule height 1.3 ex depth-1.24 ex width 0.7 ex
\kern 0.15 ex}
\def\D{D\kern-1.7 ex\vrule height .87 ex depth-0.8 ex width 0.7 ex
\kern 0.95 ex}

\begin{document}
\baselineskip 20 pt

\begin{center}

\Large{\bf Features of ion acoustic waves in collisional plasmas}

\end{center}

\vspace{0.7cm}

\begin{center}

{\bf  J. Vranjes and S. Poedts }

\vspace{0.7cm}

Centre for Plasma Astrophysics, and Leuven Mathematical Modeling and Computational Science Centre
 (LMCC), K.U.Leuven,  Celestijnenlaan 200B, 3001 Leuven,  Belgium.

\end{center}

\vspace{2cm}

{\bf Abstract:} The effects of friction on the ion acoustic (IA) wave in fully and partially ionized plasmas are studied. In a
quasi-neutral electron-ion plasma the friction between the two species cancels out exactly and the wave propagates without
any damping. If the Poisson equation is used instead of the quasi-neutrality, however, the IA wave is damped and the damping is
dispersive. In a partially ionized plasma, the collisions with the neutrals modify the IA wave beyond recognition. For a low
density of neutrals the mode is damped. Upon increasing the neutral density, the mode becomes first evanescent and then
reappears for a still larger number of neutrals. A similar behavior is obtained by varying the mode wave-length.
The explanation for this behavior is given. In an inhomogeneous plasma placed in an external magnetic field, and for
magnetized electrons and un-magnetized ions, the IA mode propagates in any direction and in this case the collisions
make it growing on the account of the energy stored in the density gradient. The growth rate is angle dependent. A
comparison with the collision-less kinetic density gradient driven IA instability is also given.

\vspace{0.7cm}


\vspace{3cm}

\pagebreak

%\vspace{0.5cm}

%\noindent{\bf I. \,\,\, INTRODUCTION}
\section{Introduction}
%\vspace{0.5cm}

Multi-component plasmas comprise different species that, in the presence of waves, may be in the state of relative
macroscopic motion. In such a situation, friction between  the  species may lead to wave damping (though not always, as
we are going to show in the forthcoming text). For example, neutrals in a weakly ionized plasma represent a barrier for
electron and ion motion in a wave field. A similar friction appears  in a fully ionized plasma when the electron and
ion components do not share the same momentum. The interaction is described by a friction force $\vec F_j= m_j n_j
\nu_{jl} (\vec v_j-\vec v_l)$ in the momentum equation for the species $j$. Momentum conservation implies that for its
counterpart $l$, $\vec F_l= m_l n_l \nu_{lj} (\vec v_l-\vec v_j)$, where  $m_j n_j \nu_{jl}= m_l n_l \nu_{lj}$. If the
two species $j$ and $l$ have a large mass difference, the friction response of the heavier component is
typically omitted as negligible in the literature. However, this may yield completely wrong results as we shall demonstrate in the
forthcoming text using the ion acoustic (IA) mode as an example. %In homogeneous plasmas these effects cause  damping of plasma modes.

In the presence of high frequency waves  $\omega \gg \Omega_i=e B_0/m_i$ in a plasma placed in an external magnetic
field $\vec B_0= B_0 \vec e_z$, ions will follow nearly straight lines regardless of the direction of the wave-number
vector $\vec k$ and the magnetic field vector. For electrons, in view of the mass difference, the opposite may hold,
$\omega \ll \Omega_e=e B_0/m_e$, hence they will behave as magnetized and their perpendicular and parallel dynamics
will be essentially different \cite{k1}. Ions can behave as un-magnetized in the perturbed state also in case of
collisions provided that  $\nu_i>\Omega_i$  even if at the same time $\Omega_i > \omega$, or for short wavelengths
$\lambda<\rho_i$, $\rho_i=v_{{\sss T} i}/\Omega_i$, $v_{{\sss T} i}^2=\kappa T_i/m_i$. In the case of an inhomogeneous
equilibrium, with a density gradient perpendicular to the magnetic field vector, in the unperturbed state the ions may
behave as un-magnetized in case of a low temperature, when   their diamagnetic drift velocity becomes negligible as
compared to electrons [for singly charged ions $v_{*i}/v_{*e}=T_i/T_e$, where $v_{*j}= \kappa T_j n_{j0}'/(q_j B_0
n_{j0})$, and $n_j'=d n_j/dx$ denotes the equilibrium density gradient]. The same holds in the presence of numerous
collisions as above, $\nu_i>\Omega_i$, when their diamagnetic effects are absent too.

In all these  situations, and  neglecting the electron polarization drift (inertia-less limit),  the wave will still
have the basic properties of the IA mode. Within the  two-fluid theory such a mode in an inhomogeneous plasma [that may be
called ion-acoustic-drift (IAD) mode] may in fact become growing \cite{k1}-\cite{v1}  in the simultaneous presence of
collisions and the mentioned equilibrium density gradient perpendicular to $\vec B_0$.

Within the  kinetic theory the mode is also growing in the presence of the same density gradient and this even without
collisions (due to purely kinetic effects), and the physics of the growth rate is similar to the standard drift wave
instability \cite{k2}. It requires that the wave frequency is below the electron diamagnetic drift frequency
$\omega_{*e}=v_{*e} k_\bot$.

On the other hand, keeping the electron inertia results in the instability of the lower-hybrid-drift (LHD) type
\cite{k3}-\cite{h1}. In some other limits the effects of the same density gradient yield growing ion plasma (Langmuir)
oscillations \cite{k3}, or growing electron-acoustic oscillations \cite{moh}.

In the present manuscript the friction force effects on the IA wave are discussed, both for fully and partially ionized
un-magnetized plasmas, and for inhomogeneous plasmas with magnetized electrons. The latter implies growing modes within
both the fluid and kinetic descriptions, and in the manuscript these two instabilities  are compared.

%\vspace{0.5cm}

%\noindent{\bf II. \,\,\, MODEL AND EQUATIONS}

\section{IA wave in fully and partially ionized collisional plasmas}
%\vspace{0.5cm}

The equations used further in this section  are the momentum equations for the ions, the electrons and the neutral particles, respectively:
\be
m_i n_i \left(\frac{\partial}{\partial t} + \vec v_i\cdot \nabla\right)\vec v_i= - e n_i \nabla \phi- \kappa T_i \nabla
n_i - m_i n_i \nu_{ie} (\vec v_i - \vec v_e) - m_i n_i \nu_{in} (\vec v_i - \vec v_n),\label{s1} \ee
\be
m_e n_e \left(\frac{\partial}{\partial t} + \vec v_e\cdot \nabla\right)\vec v_e=  e n_e \nabla \phi- \kappa T_e \nabla
n_e - m_e n_e \nu_{ei} (\vec v_e - \vec v_i) - m_e n_e \nu_{en} (\vec v_e - \vec v_n),\label{s2} \ee
and
\be
m_n n_n \left(\frac{\partial}{\partial t} + \vec v_n\cdot \nabla\right)\vec v_n=  - \kappa T_n \nabla n_n - m_n n_n
\nu_{ni} (\vec v_n - \vec v_i) - m_n n_n \nu_{ne} (\vec v_n - \vec v_e),\label{s3} \ee
and the continuity equation
\be
\frac{\partial n_j}{\partial t}+ \nabla\cdot(n_j \vec v_j)=0, \quad {j=e, i, n}. \label{s4} \ee
This  set of equations is closed either by using the quasi-neutrality or the Poisson equation. The differences between
the two cases are discussed below.

\subsection{Friction in electron-ion plasma}

The continuity equation (\ref{s4}) yields
\be
 v_{i1}=\omega n_{i1}/(k n_0), \quad  v_{e1}=\omega n_{e1}/(k n_0),
\label{c} \ee
so that the velocity difference in the friction term $v_e - v_i\equiv 0$ if the quasi-neutrality is used. The IA mode propagates without any damping. Hence, the friction force  in a fully ionized plasma in this limit
cancels out exactly even without using the momentum balance. The physical reason for this is the assumed exact   balance of
the perturbed densities: what one plasma component loses the other component receives, this is valid at every
position in the wave and no momentum is lost.

A  typical mistake seen in the literature is to take the friction force term for electrons only, in the form $m_e n_e
\nu_{ei} \vec v_e$. This comes with the excuse  of the large mass difference, so that the displacement of the much
heavier ion  fluid, caused by the electron friction is neglected. In the case of a fully ionized electron-ion plasma
this yields a false damping of the IA  mode within the quasi-neutrality limit:
\be
\omega=\pm k(c_s^2 + \vti^2)^{1/2} - \nu_{ei}/2.\label{ins} \ee
On the other hand, if the Poisson equation is used instead of the quasi-neutrality,  one obtains \cite{v1a}
\be
\omega=\pm k v_s \left(1-r_{de}^2k^2 \frac{\nu_{ie}^2r_{de}^2}{v_s^2}\right)^{1/2} - i \nu_{ie}r_{de}^2k^2.\label{puas}
\ee
Here,  we have used the momentum conservation $\nu_{ie}=m_e\nu_{ei}/m_i$ and $v_s^2= c_s^2+ \vti^2$,
$r_{de}=\vte/\omega_{pe}$. The physical reason for damping in the present case is the fact that the detailed balance
$n_{i1}=n_{e1}$ does not hold, because of the electric field which takes part for small enough wave-lengths. It can
easily be seen that for any realistic parameters the second term in the real part of the frequency in  Eq.~(\ref{puas}) is
much below unity and the mode is never evanescent. However, in partially ionized plasmas (see below) this may be
completely different.

\subsection{Friction and collisions in partially ionized plasma}

Keeping the quasi-neutrality limit, we now discuss the IA wave damping in plasmas comprising neutrals as well. In view
of the results presented above, the electron-ion friction terms in Eqs.~(\ref{s1}) and (\ref{s2}) will cancel each other
out and in a few steps one derives the following dispersion equation containing the collisions of  plasma species with neutrals
and vice versa:
\[
\omega^3+ i \omega^2 \left(\nu_{en} \frac{m_e}{m_i} + \nu_{in}\right) \left(1 + \frac{m_i}{m_n}\frac{n_0}{n_{n0}}
\right) -  k^2 c_s^2 \,\omega
\]
\be
- i k^2 c_s^2 \,\frac{m_e}{m_n} \frac{n_0}{n_{n0}} \left(\nu_{en} + \frac{m_i}{m_e} \nu_{in}\right)=0. \label{dn} \ee
In the derivation, the ion and neutral thermal terms are neglected. The ion thermal  terms would give the modified
mode frequency $\omega^2=k^2 c_s^2 (1+ T_i/T_e)$. Hence, even if $T_e=T_i$ the wave frequency is only modified by a
factor $2^{1/2}$. The neutral thermal terms are discussed further in the text.
\begin{figure}
\includegraphics[height=6cm, bb=15 15 270 220, clip=,width=.5\columnwidth]{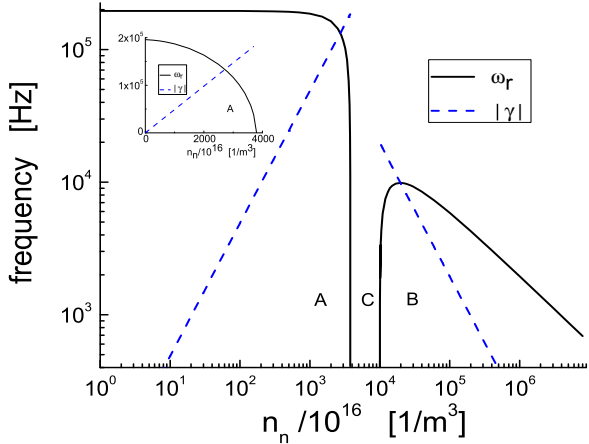}\\
\caption{\label{figs} Frequency $\omega_r$ and absolute value of the IA  mode damping $|\gamma|$ in terms of the number
density of neutrals. Details of the mode behavior in the region A are better seen in the linear scale (small figure
inside). }
\end{figure}
Note that in deriving Eq.~(\ref{dn}), the momentum conservation condition $\nu_{ie}=m_e\nu_{ei}/m_i$  is nowhere used: the
e-i and i-e friction terms exactly vanish in view of Eq.~(\ref{c}).

Equation~(\ref{dn}) is solved numerically for a plasma containing electrons, protons, and neutral hydrogen atoms using the
following set of parameters: $T_e= 4\;$eV, $n_0=10^{18}\;$m$^{-3}$, $k=10\;$m$^{-1}$, with \cite{bk} $\sigma_{en}=1.14
\cdot 10^{-19}\;$m$^{-2}$. The neutral density is varying in the interval $10^{16} - 10^{23}\;$m$^{-3}$. The ion and hydrogen
temperatures are taken $T_i=T_n=T_e/20$, satisfying the condition of their small thermal effects. This also gives
\cite{kr}, $\sigma_{in}=2.24 \cdot 10^{-18}\;$m$^{-2}$.  The results are presented in Fig.~1. The IA mode propagates in
two distinct regions A and B.

Only a limited  left part of the region A would correspond to the 'standard' IA wave behavior in a collisional plasma: the
mode is damped and the damping is proportional to the neutral number density. Hence, in this region it may be  more or
less appropriate to use the approximate expressions for the friction force, like (in the case of electrons) $F_e\simeq
m_e n_0\nu_{en} v_{e1}$. However, this domain  is very limited because in the rest of the domain the frequency drops
and the mode becomes non-propagating for $n_{n0}\geq 3.8 \cdot 10^{19}\;$m$^{-3}$ (this is the lower limit of the region C
in Fig.~1).

Increasing the neutral number density, after some critical value (in the present case this is around  $n_{n0} \simeq
10^{20}\;$m$^{-3}$) the IA mode reappears again in the region B, with a frequency starting from zero. For even larger
neutrals number densities,  the mode damping in fact vanishes completely and the wave propagates freely but with a
frequency that is many orders of magnitude below the ideal case $k c_s \simeq 196\;$kHz. This behavior can be explained
in the following manner. For a relatively small number of collisions the IA mode is weakly damped because initially
neutrals do not participate in the wave motion and do not share the same momentum. Increasing the number of neutrals,
the damping may become so strong that the wave becomes evanescent. However, for much larger collision frequencies
(i.e., for a lower ionization ratio), the tiny population of electrons and ions is still  capable of dragging neutrals
along and all three components  move together. The plasma and the neutrals become so strongly coupled that  the two
essentially different fluids participate in the electrostatic wave  together. In this regime, the stronger the
collisions are, the less wave damping there is! Yet, this a bit counter-intuitive behavior comes  with a price: the wave frequency
and the wave energy flux becomes reduced by several orders of magnitude.

\begin{figure}
\includegraphics[height=6cm, bb=15 15 270 220, clip=,width=.5\columnwidth]{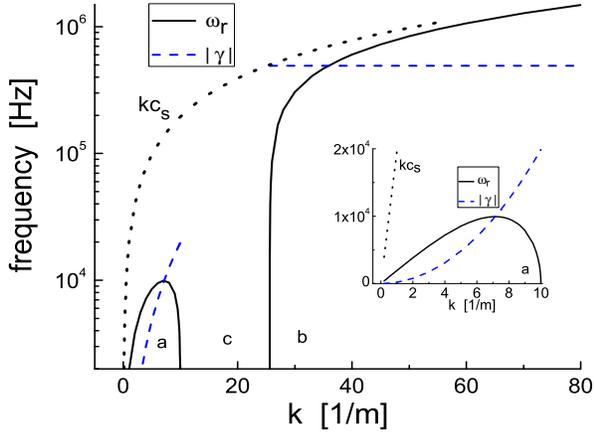}\\
\caption{\label{figk} Frequency $\omega_r$ and absolute value of the IA  mode damping $|\gamma|$ in terms of the wave
number. The line $k c_s$ shows part of the graph of the ideal mode. Details of the domain $a$ are better  seen in the
linear scale (small figure inside). }
\end{figure}

Similar effects may be expected  by varying the wave-length. The previous role of the varying density of neutrals is
now replaced by the  the ratio of the mean free path of a species $\lambda_{fj}=v_{{\sss T} j}/\nu_j$ (with respect to
their collision with neutrals) and the wavelength. This ratio now determines the  coupling between the plasma and
the neutrals. The mode behavior is directly numerically checked by fixing $n_{n0}=10^{20}\;$m$^{-3}$,
$n_0=10^{18}\;$m$^{-3}$, and for other parameters same as above. For these parameters we have $\lambda_{fe}=v_{{\sss T}
e}/\nu_{en}=0.09$ m, and $\lambda_{fi}=v_{{\sss T} i}/\nu_{in}=0.004$. The numerical results are presented in Fig.~2
for $k$ varying in the interval $0.2 - 80\;$m$^{-1}$. The mode vanishes in the interval $c$, between
$k\simeq 10\;$m$^{-1}$ and $k\simeq 25.6\;$m$^{-1}$. The explanation is similar as before. Note that for $k=0.2\;$m$^{-1}$ (in the
region $a$)  we have $\omega_r\simeq 390\;$Hz, and this is about one order below $k c_s$. Compared to the mode behavior
in Fig.~1, this implies  that the mode in the present domain $a$  is  in the regime equivalent  to the domain B from
Fig.~1; here, in Fig.~2, these large wave-lengths imply  well coupled plasma-neutrals, where the frequency is reduced and
the damping is small. The region $a$ is also given separately in linear scale together with the dotted line describing the
ideal mode $k c_s$. Clearly, in general the realistic behavior of the wave is beyond recognition and completely
different as compared to the ideal case.
\begin{figure}
\includegraphics[height=6cm, bb=25 10 655 530, clip=,width=.5\columnwidth]{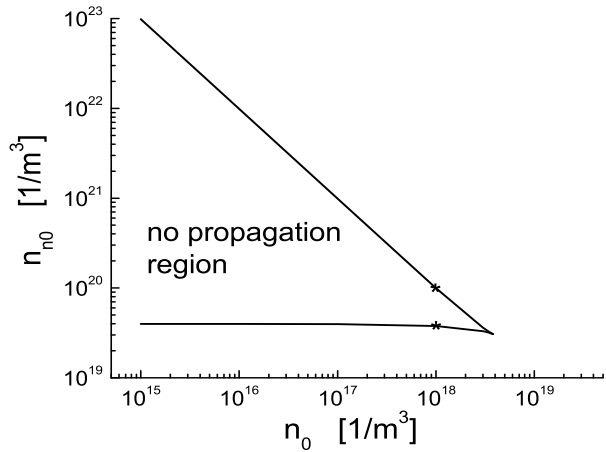}
\caption{\label{figl} The two lines give the lower and upper values of the neutrals' density $n_{n0}$ between which,
for the given plasma density $n_0$, the  IA mode does not propagate. }
\end{figure}

After checking for various sets of plasma densities, it appears that the evanescence region reduces and vanishes  for
larger plasma densities $n_0$. This is presented in Fig.~3 for the same parameters as above, by  taking $k=10\;$m$^{-1}$,
but for a varying plasma density $n_0$. The two lines represent boundary values of the number densities of
neutrals, for the given plasma density,  at which the IA mode vanishes;  for the neutrals densities between the two
lines the IA mode does not propagate. The symbols $*$ on the two lines denote the boundaries of the region $C$ from
Fig.~1. It is seen that for the given case the IA mode propagates without evanescence for the plasma densities above
$n_0=3.8 \cdot 10^{18}\;$m$^{-3}$. Physical reason for a larger non-propagating domain for low plasma density is
obvious, namely the tiny plasma population is less efficient in inducing a synchronous motion of neutrals. In the other
limit, the opposite happens and the forbidden region eventually vanishes.

\begin{figure}
\includegraphics[height=6cm, bb=40 14 655 525, clip=,width=.5\columnwidth]{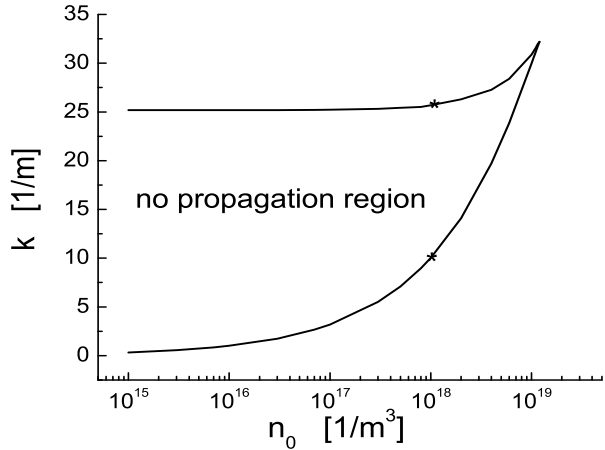}
\caption{\label{fig11} Values of the wave-number, in terms of the plasma density,  for which the IA wave becomes
evanescent. In the region between the lines the mode does not propagate.  }
\end{figure}

A similar check is done by varying the wave-number and the plasma density, and the result is presented in Fig.~4 for a
fixed $n_{n0}=10^{20}\;$m$^{-3}$. The lines represent the values $(n_0, k)$ at which the IA wave becomes evanescent.
There can be no wave in the region between the lines. On the other hand, there is no evanescence for the plasma density
above $n_0=1.2 \cdot 10^{19}\;$m$^{-3}$. Here $*$ denote the boundaries of the region $c$ from Fig.~2.

All these results  clearly indicate that  in practical measurements in laboratory and space plasmas, the IA mode
can hardly be detected and recognized as the IA mode unless collisions are correctly  taken into account (using full
friction terms), and the mode is sought in the corresponding domain which follows from our Eq.~(\ref{dn}).

\subsection{Thermal effects of neutrals}

\begin{figure}
\includegraphics[height=6cm, bb=15 15 280 225, clip=,width=.5\columnwidth]{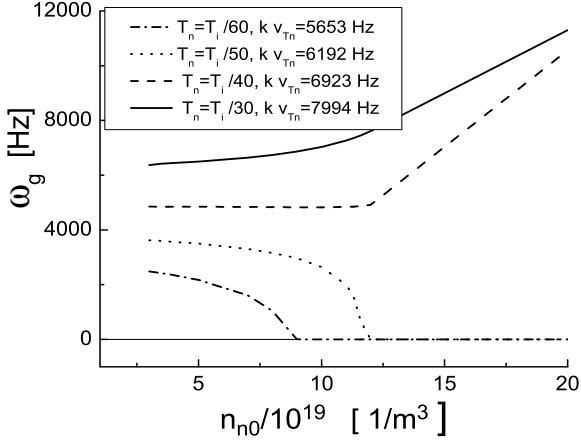}
\caption{\label{gg} The real part of the frequency of damped gas thermal mode in terms of the number density of
neutrals and for several temperatures of the neutrals gas.  }
\end{figure}

Keeping the pressure terms for ions and neutrals yields the following dispersion equation
\[
\omega^4+ i \omega^3 \left(\nu_{in} + \nu_{en} \frac{m_e}{m_i}\right) \left(1+ \frac{m_i}{m_n}
\frac{n_0}{n_{n0}}\right) - k^2 \left(v_s^2 + \vtn^2\right) \omega^2
\]
\be
- i \omega k^2 \left[\frac{n_0}{n_{n0}} \frac{m_i}{m_n}v_s^2 \left(\nu_{in} + \nu_{en} \frac{m_e}{m_i}\right) +
\nu_{in} \vtn^2\right] + k^4 \vtn^2 v_s^2=0. \label{g} \ee
Here, $v_s^2=c_s^2 + \vti^2$. Without collisions, this yields two independent modes, viz.\ the ion-acoustic mode and the
gas thermal (GT) mode,  $(\omega^2- k^2 \vtn^2)(\omega^2- k^2 v_s^2)=0$. The collisions couple the two modes, and in order to compare with
the previous cases we solve Eq.~(\ref{g}) for $k=10\;$m$^{-1}$, $n_0=10^{18}\;$m$^{-3}$, $T_e=4\;$eV, $T_i=T_e/20$,  and in
terms of the density and temperature of neutrals. For a low thermal contribution  of neutrals  (i.e., a low neutral
temperature, or/and heavy neutral atoms)  the previous results remain valid.  Larger values of  $\vtn$ introduce new
effects, this is checked by varying the temperature $T_n$. The ion thermal terms do not  make much difference, as
explained earlier. The real part of the frequency $\omega_g$ of the  gas thermal mode is presented in Fig.~\ref{gg},
and this only in a limited region that  includes the  evanescence area $C$ from Fig.~\ref{figs}. The damping is not
presented but the mode is in fact heavily damped.

The explanation of the figure is as follows. The starting solution for $T_n=0$ is in fact the line $\omega_g=0$, and
this case would correspond to the the IA mode from  Fig.~\ref{figs}. For some finite $T_n$ there appears the GT mode.
For a low gas temperature the mode becomes evanescent for a higher density of neutrals (the dot  and dash-dot  lines in
Fig.~\ref{gg}). This evanescence is accompanied with the previously discussed evanescence and re-appearance of the IA
mode (described earlier and no need to be presented  here again). However, for still larger $T_n$, the IA and GT modes
become indistinguishable and propagate as one single mode. This is presented by the two upper (the full and
dashed) lines in Fig.~\ref{gg}, that go up for large enough $n_{n0}$. Also  given are the corresponding ideal values $k
\vtn$ that appear to be  much above the actual wave frequency $\omega_g$ in such a  collisional plasma, but this
remains so only until the  neutral density $n_{n0}$ exceeds  some critical value. After that the wave in fact behaves
as less and less collisional and the wave frequency is increased.

\section{IA wave instability  in inhomogeneous  partially ionized plasma}

\subsection{Fluid description in collisional plasma}

In the previous text, collisions were shown to yield damping of the IA mode. However, if the plasma is inhomogeneous,
implying the presence of source of free energy in the system, a drift-type instability of the IA wave may develop if there is
a magnetic field
$\vec B_0 = B_0 \vec e_z$ present, and the electrons (ions) are magnetized (un-magnetized). The magnetic field introduces a
difference in the parallel and perpendicular dynamics of the magnetized species so that the continuity condition in this
case can be written as
\be
\frac{\partial n_{j1}}{\partial t} + n_{j0} \nabla\cdot \vec v_{j1} + \vec v_{j1}\cdot \nabla n_{j0}=0. \label{e2} \ee
Here, $\nabla\equiv \nabla_\bot + \nabla_z$. For  the {\em un-magnetized} species the direction of the wave plays no role
so that $\nabla\rightarrow i \vec k$, $k^2=k_y^2+ k_z^2$. On the other hand, for the equilibrium gradient along the
$x$-axis and for perturbations of the form $\sim f(x)\exp(-i \omega t + i k_y y + i k_z z)$, where $|(d f/dx)/f|, |(d
n_{j0}/dx)/n_0|\ll k_y$, we apply a local approximation, and for ions the last term in Eq.~(\ref{e2}) vanishes because of the
assumed geometry. The ions' dynamics is basically the same as in the previous sections.

The electron momentum equation (\ref{s2}) will now include the Lorentz force term $- e n_e \vec v_e\times \vec B$.
Repeating the derivation from Ref.~\cite{v1}, the total perpendicular electron velocity can be written as
 \be
  v_{e\bot}= \frac{1}{1+ \nu_{en}^2 \alpha^2/\Omega_e^2}\left[\frac{1}{B_0}\vec
e_z\times \nabla_\bot  \phi +\frac{\nu_{en} \alpha}{\Omega_e} \frac{\nabla_\bot \phi}{B_0} - \frac{v_{\sss{T}
e}^2\nu_{en} \alpha}{\Omega_e^2} \frac{\nabla_\bot n_e}{n_e}  -  \frac{v_{\sss{T} e}^2}{\Omega_e} \vec e_z\times
\frac{\nabla_\bot n_e}{n_e}\right]. \label{e5} \ee
In the direction along the magnetic field vector, the perturbed electron  velocity is
 \be
v_{ez1}=\frac{i k_z \vte^2}{\nu_{en}} \frac{\omega^2 + \nu_{ne}^2}{\omega^2 - i \nu_{ne} \omega} \left(\frac{e
\phi_1}{\kappa T_e} - \frac{n_{e1}}{n_0} \right). \label{e6} \ee
Here, $\alpha=\omega/(\omega + i \nu_{ne})$, and for magnetized electrons,  $|\nu_{en}^2 \alpha^2/\Omega_e^2|\ll 1$ in
the denominator in Eq.~(\ref{e5}).  Using these equations in the continuity condition (\ref{e2}) for electrons one
obtains
 \be
\frac{n_{e1}}{n_0}= \frac{\omega_{*e} + i D_p + i D_z (\omega^2 + \nu_{ne}^2)/(\omega^2- i \nu_{ne} \omega)}{ \omega +
i D_p + i D_z (\omega^2 + \nu_{ne}^2)/(\omega^2- i \nu_{ne} \omega)} \frac{e \phi_1}{\kappa T_e}, \label{e7} \ee
\[
  D_p= \nu_{en}\alpha k_y^2 \rho_e^2, \quad D_z= k_z^2 \vte^2/\nu_{en}, \quad \rho_e=\vte/\Omega_e.
 \]
The term $D_p$ describes the effects of collisions on the electron perpendicular dynamics and is usually omitted in the
literature. However, as shown in a recent study \cite{v1}, it can strongly modify the growth rate of the drift and
IA-drift wave instability in the limit of small parallel wave-number $k_z$.

Neglecting the neutral dynamics  is equivalent to setting $\nu_{ne}=0$. This yields $\alpha=1$, and Eq.~(\ref{e7}) becomes
 identical to the corresponding expression in Refs.~\cite{mih,v5}. For  a negligible $D_p$, Eq.~(\ref{e7}) becomes the
  same as the corresponding equation from Ref.~\cite{v3}.
For negligible ion thermal effects, the final dispersion equation reads
\be
 \frac{k^2 c_s^2 }{\omega^2}  =   \frac{\omega_{*e} + i D_p + i D_z (\omega^2 + \nu_{ne}^2)/(\omega^2- i \nu_{ne} \omega)}{
\omega  + i D_p + i D_z (\omega^2 + \nu_{ne}^2)/(\omega^2- i \nu_{ne} \omega)}. \label{e9a} \ee
Equation~(\ref{e9a}) can be solved numerically keeping in mind a number of conditions used in their derivations, like smallness of
the plasma beta to remain in electrostatic limit, smallness of the parallel phase velocity as compared to the electron thermal
speed because of the massless electrons limit, also the ratio $D_p/D_z$ should be kept not too big or too small in
order to have the assumed effects of electron collisions in perpendicular direction. We plan to compare this
collisional instability with the kinetic instability due to  the presence of the density gradient. Therefore, the wave
frequency should be below the electron diamagnetic frequency etc.

\begin{figure}
\includegraphics[height=7cm, bb=5 5 280 220, clip=,width=.6\columnwidth]{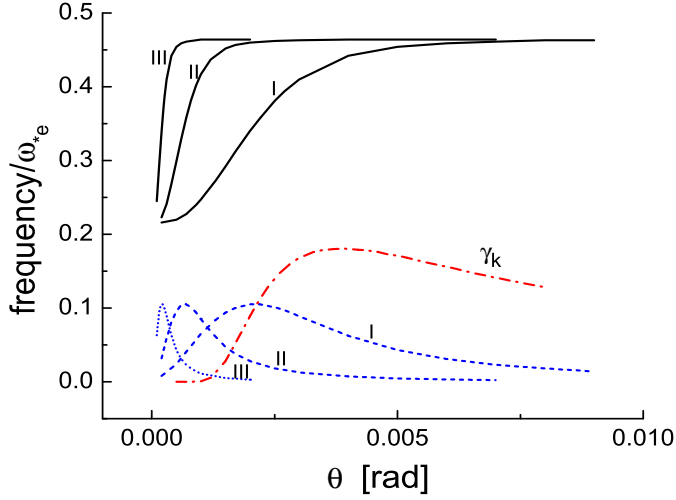}
\caption{\label{fig1a} Real part of the frequency from Eq.~(\ref{e9a}) (full lines) and the corresponding growth rates
(dashed lines), both normalized to the electron diamagnetic drift frequency,  for three values of neutral number
density. The lines I, II, III correspond (respectively) to $n_{n0}=10^{19}, \, 10^{18}, \,10^{17}\;$m$^{-3}$. The line
$\gamma_k$ is the kinetic growth-rate from Eq.~(\ref{kg}) (for the same  parameters as line II). }
\end{figure}

\begin{figure}
\includegraphics[height=7cm, bb=5 5 280 220, clip=,width=.6\columnwidth]{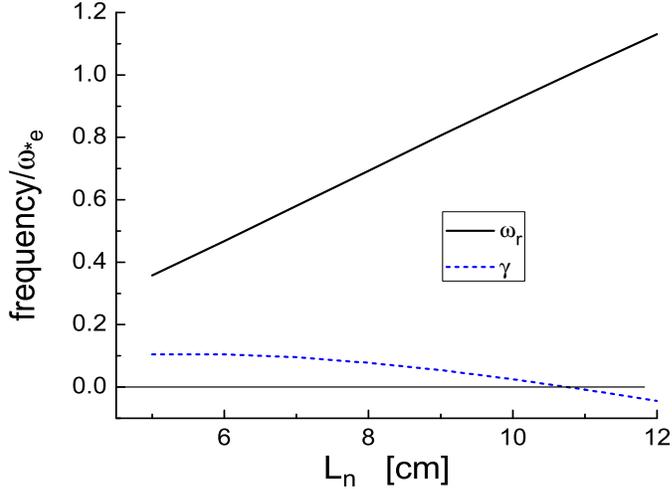}
\caption{\label{fig1b} The real and imaginary parts of the frequency for the line II from Fig.~\ref{fig1a},  in terms
of the characteristic density inhomogeneity scale length $L_n=(dn_0/dx)^{-1}$, and for the angle $\theta$ at  the
maximum on Fig.~\ref{fig1a}.}
\end{figure}

We solve Eq.~(\ref{e9a}) for an electron-argon  plasma in the presence of parental argon atoms. As an example we take
$T_e= 4\;$eV, $T_i=T_n=T_e/30$, $n_0=10^{15}\;$m$^{-3}$, $B_0=1.2\cdot 10^{-2}\;$T, $k=500\;$m$^{-1}$, $L_n =0.05\;$m, and take
several values for the density of neutrals. The result in terms of the angle of the propagation
$\theta=\arctan(k_z/k_y)$  is presented in Fig.~5. The three lines (full for the real part of the frequency, and dashed
for the growth rates) are for $n_{n0}=10^{19}, \, 10^{18}, \,10^{17}\;$m$^{-3}$. It is seen that i) the instability is
angle dependent and there exists an  angle of preference and an instability window in terms of $\theta$ within  which
the mode is most easily excited, ii) this angle of preference is shifted towards smaller values for lower values of the
neutral density, and iii) in the same time the instability window becomes considerably reduced. This shows an
interesting  possibility of launching the IA-drift wave in a certain direction by simply varying the pressure of the
neutral gas.

Varying the density scale length  $L_n=(dn_0/dx)^{-1}$ the wave frequency may become above $\omega_{*e}$ and in this
case the instability vanishes. As an example, this is demonstrated in Fig.~\ref{fig1b} for the parameters corresponding
to the line II from Fig.~\ref{fig1a} and for  the angle $\theta$  at the maximum growth rate. The growth rate changes
the sign for $\omega\simeq \omega_{*e}$.

\subsection{Comparison with collision-less kinetic gradient-driven IA wave instability}

Keeping the same model of magnetized (un-magnetized) electrons (ions), within the kinetic theory the perturbed number
density for electrons can be written as \cite{w}
\be
\frac{n_{e1}}{n_0}= \frac{e \phi_1}{\kappa T_e} \left\{1+ i \left(\frac{\pi}{2}\right)^{1/2}
\frac{\omega-\omega_{*e}}{k_z \vte} \exp\left[-\omega^2/(2 k_z^2 \vte^2)\right]\right\}. \label{ew} \ee
In the derivation of Eq.~(\ref{ew}) the electron Larmor radius corrections are neglected in terms of the type $I_n(b)
\exp(-b)$, $b=k_\bot^2 \rho_e^2$, where $I_n$ denotes the modified Bessel function of the first kind, order $n$, and
only $n=0$ terms are kept  for the present case of frequencies much below the gyro-frequency.

The ion number density can be calculated using the kinetic description for un-magnetized species, the derivation is
straight-forward and it yields \cite{v6}
\be
\frac{n_{i1}}{n_{i0}}= - \frac{e \phi_1}{m_iv_{{\sss T} i}^2} \left[ 1- J_+ \left(\frac{\omega_i}{k v_{{\sss
T}i}}\right)\right]. \label{e11} \ee
Here, $J(\eta)= [\eta/(2 \pi)^{1/2}] \int_c d \zeta\exp(- \zeta^2/2)/(\eta-\zeta)$ is the plasma dispersion function,
and $\zeta=v/v_{{\sss T} i}$. In the case $|\eta|\gg 1$, and assuming $|Re(\eta)|\gg Im(\eta)$, an  expansion is used
for  $J(\eta)$. This together with the quasi-neutrality yields the kinetic  dispersion equation for the IA-drift wave:
\[
\Delta(\omega, k) \equiv 1-\frac{k^2 c_s^2}{\omega^2} - \frac{3 k^4 \vti^2 c_s^2}{\omega^4}
\]
\be
+ i (\pi/2)^{1/2} \left\{\frac{\omega-\omega_{*e}}{k_z \vte} \exp\left[-\omega^2/(2 k_z^2 \vte^2)\right]
+\frac{T_e}{T_i} \frac{\omega}{k \vti} \exp\left[-\omega^2/(2 k^2 \vti^2)\right]  \right\}. \label{ek} \ee
The real part of Eq.~(\ref{ek}) yields the spectrum
\be
\omega_k^2=\frac{k^2 c_s^2}{2} \left[1+ \left(1+ 12T_i/T_e\right)^{1/2}\right]. \label{s} \ee
The kinetic growth rate is given by
\[
\gamma_k \simeq -Im \Delta/(\partial Re \Delta/\partial \omega)=-\frac{(\pi/2)^{1/2} \omega_k^3}{ 2 k^2 c_s^2} \times
\]
\be
\times \left\{\frac{\omega_k-\omega_{*e}}{k_z \vte} \exp\left[-\omega_k^2/(2 k_z^2 \vte^2)\right] +\frac{T_e}{T_i}
\frac{\omega_k}{k \vti} \exp\left[-\omega_k^2/(2 k^2 \vti^2)\right]  \right\}.\label{kg} \ee
Here, the index $k$ is used to denote kinetic expressions.  The electron contribution in Eq.~(\ref{kg}) yields a kinetic
instability provided that $\omega_k< \omega_{*e}$.

Equation~(\ref{kg}) is solved numerically and compared with the growth rate obtained from the collisional IA-drift mode
(\ref{dn}). For a fixed $k=500\;$m$^{-1}$ as in Figs.~\ref{fig1a} and \ref{fig1b}, the normalized frequency
$\omega_k/\omega_{*e}=0.485$, and the result for the growth rate is presented by the line $\gamma_k$ in
Fig.~\ref{fig1a} for the parameters corresponding to the line II from the fluid analysis (i.e., for $n_{n0}= 10^{18}\;$m$^{-3}$).
The larger kinetic growth rate appears also to be  angle dependent, yet with a much wider instability window
as compared to the collisional gradient driven instability obtained from the fluid theory.

\section{Summary}

The analysis of  the ion acoustic wave presented here shows the importance of  collisions in describing the wave
behavior. Without a proper analytical description, the identification of the  mode in the laboratory and space
observations may be rather difficult because one might fruitlessly search for the wave in a very inappropriate domain,
as can be concluded from the graphs presented here, and in particular from Fig.~2. Not only the wave frequency may
become orders of magnitude below an expected ideal value, but also the mode may completely vanish. A similar analysis
of the effects of collisions may be performed  for other plasma modes as well, like the Alfv\'{e}n wave etc, as
predicted long ago in classic Ref.~\cite{tan}. The impression is that these effects are frequently overlooked in the
literature, hence the necessity for the quantitative analysis given in the present work that can be used as a  good
starting point for an eventual experimental check of the wave behavior in collisional plasmas. Particularly interesting
for experimental investigations may be the angle dependent mode behavior given in Sec.~3, where it is shown that the
strongly growing mode may be expected within a given narrow instability window in terms of the angle of propagation.
Comparison with the kinetic theory shows a less pronounced angle dependent peak, yet this kinetic effect can
effectively be smeared out in the presence of numerous collisions, that are known to reduce kinetic effects in any
case, and the sharp angle dependence that follow from pure fluid effects should become experimentally detectable.

\vspace{1cm}

\paragraph{Acknowledgements:}
The  results presented here  are  obtained in the framework of the projects G.0304.07 (FWO-Vlaanderen), C~90347
(Prodex),  GOA/2009-009 (K.U.Leuven). Financial support by the European Commission through the SOLAIRE Network
(MTRN-CT-2006-035484) is gratefully acknowledged.

\pagebreak


\begin{thebibliography}{99}
\bibitem{k1} N. A. Krall, in   {\it Advances in Plasma Physics}, ed. A Simon and W. B. Thompson  (Interscience, New York, 1968), vol. 1, p.
195.
\bibitem{mih} A. B. Mikhailovskii,  {\it Theory of Plasma Instabilities}  (Consultants Bureau,  New York, 1974), vol. 2,
p. 192.
\bibitem{v1} J. Vranjes and S. Poedts,  Phys. Plasmas {\bf 16}, 022101 (2009).
\bibitem{k2} N. A. Krall and D. Book,  Phys. Rev. Lett. {\bf 23}, 574 (1969); {\it ibid.}, Phys. Fluids {\bf 12}, 347
(1969).

\bibitem{k3} N. A. Krall and P. C. Liewer,  Phys. Rev. A {\bf 4}, 2094 (1971).
\bibitem{moh} M. Mohan and M. Y. Yu,  J. Plasma Phys. {\bf 29}, 127 (1983).
\bibitem{bos} M. Bose and S. Guha,   Phys. Scripta {\bf 34}, 63 (1986).
\bibitem{h1} J. D. Huba, A. B. Hassam, and D. Winske,   Phys. Fluids B {\bf 2}, 1676 (1990).
\bibitem{v1a} J. Vranjes, M. Kono, S. Poedts, and M. Y. Tanaka,  Phys. Plasmas {\bf 15}, 092107 (2008).
\bibitem{bk} B. Bedersen and L. J. Kiefer, Rev. Mod. Phys. {\bf 43}, 601 (1971).

%V. E. Fortov and I. T. Iakubov,    {\em The Physics of Non-Ideal Plasma} (World Scientific, Singapore, 2000) p. 380.
\bibitem{kr} P. S. Krstic and D. R. Schultz,  J. Phys. B: Mol. Opt. Phys. {\bf 32}, 3485  (1999).
%\bibitem{v1} J. Vranjes and S. Poedts,   Phys. Plasmas {\bf 16}, 022101 (2009).
\bibitem{v5} J. Vranjes, D. Petrovic, B. P. pandey,   and S. Poedts,   Phys. Plasmas {\bf 15}, 072104 (2008).
\bibitem{v3} J. Vranjes   and S. Poedts,   Phys. Plasmas {\bf 15}, 034504 (2008).
\bibitem{w} J.  Weiland,   {\em Collective Modes in Inhomogeneous
Plasmas} (Institute of Physics Pub., Bristol, 2000) p. 59.
\bibitem{v6} J. Vranjes   and S. Poedts,   Eur. Phys. J. D {\bf 40}, 257 (2006).
\bibitem{tan} B. S. Tanenbaum and D. Mintzer, Phys. Fluids {\bf 5}, 1226 (1962).


\end{thebibliography}
\end{document}